# International Collaboration in Science:

# The Global Map and the Network

Loet Leydesdorff,[a]* Caroline Wagner,[b] Han Woo Park,[c] and Jonathan Adams[d]

## Abstract

The network of international co-authorship relations has been dominated by certain European nations and the USA, but this network is rapidly expanding at the global level. Between 40 and 50 countries appear in the center of the international network in 2011, and almost all (201) nations are nowadays involved in international collaboration. In this brief communication, we present both a global map with the functionality of a Google Map (zooming, etc.) and network maps based on normalized relations. These maps reveal complementary aspects of the network. International collaboration in the generation of knowledge claims (that is, the context of discovery) changes the structural layering of the sciences. Previously, validation was at the global level and discovery more dependent on local contexts. This changing relationship between the geographical and intellectual dimensions of the sciences has implications for national science policies.

**Keywords:** co-authorship, map, global, network, internationalization, country, European Union

[a] Amsterdam School of Communication Research (ASCoR), University of Amsterdam, Kloveniersburgwal 48, 1012 CX Amsterdam, The Netherlands; loet@leydesdorff.net ; http://www.leydesdorff.net; * corresponding author
[b] John Glenn School of Public Affairs, The Ohio State University, Columbus, OH 43210, USA; cswagner@mac.com .
[c] Department of Media & Communication, YeungNam University, 214-1, Dae-dong, Gyeongsan-si, Gyeongsangbuk-do, South Korea, 712-749; hanpark@ynu.ac.kr.
[d] *Evidence* Thomson Reuters, 103 Clarendon Road, Leeds, UK LS2 9DF; jonathan.adams@thomsonreuters.com

## Introduction

International collaboration in science has increased rapidly in recent decades (NSB, 2012, at pp. 5-37 ff.). One driver of this development has been the efforts of the European Commission to stimulate collaboration within the European Union across sectors and nations (Glänzel & Schlemmer, 2007); but this development also self-organizes at the global level of the United States and other advanced industrial nations for reasons driven by the demands of science. Mass data storage, scientific "grand challenges," electronic communications (Barjak *et al.*, 2013), and less expensive travel may also be among the drivers and facilitators (Adams, 2012). Some governments of notably smaller nations (e.g., South Korea; cf. Kwon *et al.*, 2012) invest purposefully in the stimulation of "internationalization."

The implications are profound for governance of the sciences as well as knowledge creation, since the context of discovery is no longer local or institutionalized disciplinarily in university departments (Gibbons *et al.*, 1994). For example, Kwon *et al.* (2012) found that international co-authorship relations in South Korea have considerably been increased since the late 1900s while national collaborations has steadily declined. Zhou & Glänzel (2010) and Leydesdorff & Sun (2009) showed that the national publication systems of both China and Japan have gained a synergy from foreign co-authorship relationships. But it is still debatable whether international collaboration is positively associated with the quality of scientific outputs in terms of citation impact when controlling for countries and fields (Persson *et al.*, 2004; Persson, 2010).

Coauthorship relations are a most formal indicator of international collaboration. Scientific collaborations may lead to a number of outcomes of which a co-authored paper is only one (Laudel, 2002; Katz & Martin, 1997). However, from the perspective of the development of the sciences as publication systems, the submission of manuscripts containing new knowledge claims is the crucial outcome. Furthermore, we acknowledge that coauthorship in itself does not imply that collaboration has occurred (Woolgar, 1976). It represents outcomes that the listed authors jointly view as notable, which serves as a socio-cognitive filter on the multitude of relations in the social context of discovery (Melin and Persson, 1996).

No researcher unnecessarily shares authorship and thus collaborative publication can be considered as an indicator of esteem and shared intellectual contributions. From a methodological perspective, coauthorship counts have the advantage of being reproducible over time and traceable year-on-year. The network of coauthorship relations offers a perspective on the ranks and positions of countries which provides an alternative to ranking shares of publications and citations.

Wagner & Leydesdorff (2005) suggested that international collaboration tends to free scholars from local constraints such as funding by national government agencies and social (linguistic, cultural) contexts having a direct impact on intellectual agendas. Wagner (2008) hypothesized the emerging layer of international collaborations as a “new invisible college” (cf. Crane, 1972). Leydesdorff & Wagner (2008), however, noted the formation of a central group of highly functioning nations while other nations tend to remain peripheral, possibly reinforcing a core-periphery model originally proposed by Ben-David (1971; cf. Choi, 2012; Schott, 1991). Using network statistics and cosine-normalization, these authors identified a core set of 12 European nations, the USA, and Russia in both 2005 and 2006, whereas other countries (e.g., Canada, China, and Portugal) could be considered at that time as peripheral. Language can also be associated with disadvantages in terms of access, particularly in the humanities and the social sciences (Larivière *et al*., 2006), since most bibliographic databases are focused on English literature.

In this study, we present an update of the network for 2011 using the most recently available edition of the *Science Citation Index* (SCI). As previously, we use the DVD version of this index containing 3,744 journals. This selection from the 8,336 journals covered by the *Science Citation Index-Expanded* (SCI-E) at the Web-of-Science (WoS), can be considered as the most policy-relevant because it includes the most elite and highly cited of the refereed journals. The same data is, for example, used for the *Science and Engineering Indicators* series of the National Science Board of the USA (NSB, 2012, at pp. 5-37 ff.), which also includes an index of international collaborations for 2010 in tabular format. Our study provides complementary network and visualization techniques that enable the user to envisage the effects of this globalization and additionally to zoom in to specific regions and/or networks of specific nations (Wagner *et al*., in preparation).

**Methods and materials**

One of us downloaded the entire set of the DVD-version of the *Science Citation Index* 2011; this data was then brought under the control of relational database management (in the dbf-format using Flagship v7). The data contains 1,042,654 papers of which 778,988 fulfill two conditions: (*i*) a country address is provided[1] and (*ii*) they are part of the subsets of (719,327; 69.0%) articles, (37,685; 3.6%) reviews, and (29,989; 2.9%) letters. Ephemera (such as editorial materials and meeting abstracts) were not included in our analysis. In the download, 254 country names could be distinguished, of which 201 valid entities were used as variables to the (778,988) documents as units of analysis. More than 99% of this data is in English!

An asymmetrical matrix of documents versus countries was saved as a systems file in SPSS (v20) for generating, among other things, a cosine matrix between the 201 variables (countries). UCINet (v6.28) was used to generate a symmetrical co-authorship matrix among countries (after changing all values to binary) where a record with three addresses in country A and two addresses in country B is counted as a single relation between these two countries. (An affiliations routine in social-network analysis would otherwise count this as 3 x 2 = 6 relations.) Additionally, the papers were fractionally counted: fractional counting means attribution of each address to a paper proportional to the number of addresses provided in the byline of the article. For example, if two of the three addresses are in country A, the paper is attributed for 2/3rd to this country and for 1/3rd to country B.

Among these papers 193,216 (that is, 24.8% of the 778,988 documents under study) were internationally coauthored with 825,664 addresses (39.3% of 2,101,384). Note that these numbers are somewhat greater than but not substantially different from 2005, with 23.3% of the papers internationally coauthored carrying 36.5% of the addresses (Wagner & Leydesdorff, 2008, at p. 319).

[1] Addresses in England, Scotland, Wales, and Northern Ireland were recoded as "UK".

Both the co-occurrence matrix and the cosine-normalized matrix were further processed in Pajek[2] and VOSViewer[3] for the network analysis and visualization, respectively. Using the GPS Visualizer at http://www.gpsvisualizer.com/map_input?form=data and thresholds of minimally 500 fractionally-counted papers for each country and 500 international relations for each link, a global map of international collaborations was generated; this map is available at http://www.leydesdorff.net/intcoll/intcoll.htm. The links were not weighted according to the number of coauthorship relations because this would overload the visual. Instead, a legend is inserted, and in the interactive format one can click on each link to obtain the number of collaborations in a descriptor of the Google Map.

## Results

### *a. The geographical map*

The global map of science at http://www.leydesdorff.net/intcoll/intcoll.htm provides users with an overview and all the functionalities of a Google Map, such as zooming and tagging. For example, one can click on each node and obtain the number of internationally coauthored papers based on fractionally counted papers in the set of 778,988. The links are all counted as unity (as explained above). Links can also be clicked or read from the legend. The nodes are sized proportionally to the logarithm of the number of papers.

---

[2] Pajek is a network visualization and analysis program freely available for non-commercial usage at http://pajek.imfm.si/doku.php?id=download .
[3] VOSViewer is a program for network visualization freely available at http://www.vosviewer.com .

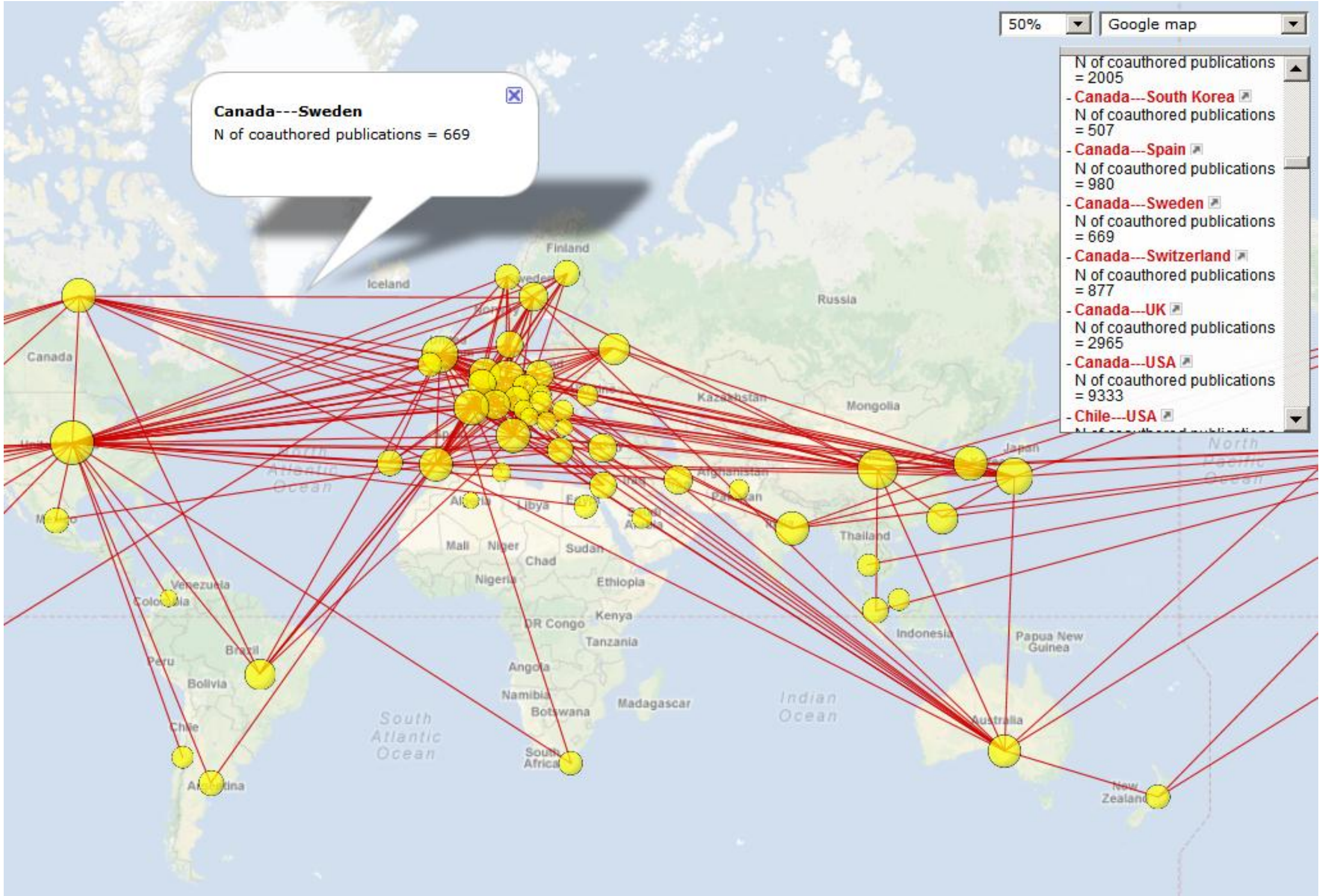

**Figure 1**: map of international collaborations; the size of each node is proportional to the logarithm of the number of fractionally counted papers. Only countries with more than 500 papers are included. The descriptors of the nodes contain the number of fractionally counted papers. (Available at http://www.leydesdorff.net/intcoll/intcoll.htm.)

As Figure 1 shows, 440 of the 12,339 links between nations surpass the threshold of more than 500 co-authorship relations (of the [201 * 200 / 2] = 20,100 possible links); 53 nations are involved. Thus, international collaboration is heavily concentrated. As an example, the link between Canada-Sweden is highlighted in the descriptor and centered in the legend table to Figure 1. Visual inspection of the map shows that from the sub-Saharan countries only South

Africa contributes, and within Latin America participation is limited to Brazil, Argentina, Chile, Venezuela, and Mexico (Wagner & Wong, 2012).

The network among EU nations is very dense. Integration makes the USA appear to operate as another member state of the EU. (One can zoom in using Google Maps online.) However, China has now become the first partner of the USA in terms of international co-authorship (that is, 12,450 integer-counted papers against 11,337 coauthored with an address in the UK). Recent accession countries (e.g., Romania and Bulgaria) are not connected given the threshold of 500 links, and smaller EU nations such as Cyprus ($N$ = 406) and Malta ($N$ = 70) are excluded because of the size restriction on the nodes. In fact, the EU-27 is not even a complete network in this (2011) set with at least one document coauthored between every country pair because of Malta and Luxembourg.

*b. The network map*

In a network map, two agents are positioned close to each other if they communicate intensively, but not on the basis of fixed (e.g., geographical) coordinates. From this different perspective, the USA would be more closely related to most EU countries than, for example, nearby Serbia. One has options to optimize the network visualization based on individual relations using a spring-based layout like that provided by Kamada & Kawai (1989)—available in Pajek—or in terms of the distributions of relations. Two nations may not relate intensively, but may share a common pattern of relations with third parties. The cosine-normalization for size captures this comparison among distributions because the cosine can also be considered as a proximity measure (comparable to the Pearson correlation, but without the reference to the mean; cf. Ahlgren *et al.*, 2003).

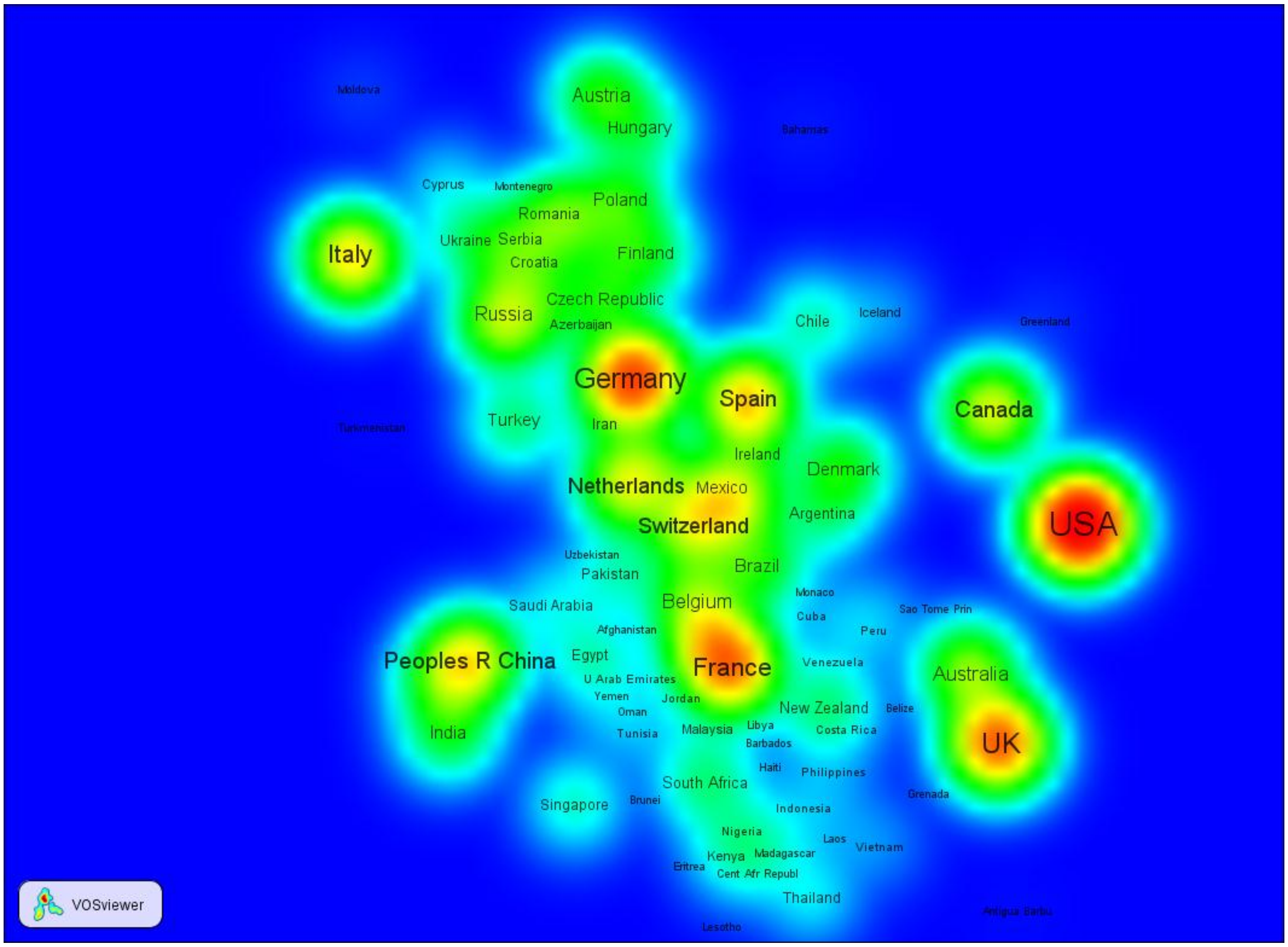

**Figure 2**: Global map based on the cosine-normalized network of coauthorship relations among 190 nations; VOSViewer used for visualization. This map can be web-started at http://www.vosviewer.com/vosviewer.php?map=http://www.leydesdorff.net/intcoll/intcoll.txt&view=2&zoom_level=1.8 .

Figure 2 shows the network of international coauthorship relations among 190 countries. Some smaller nations (such as Kosovo, Gibraltar, and the Netherlands Antilles) were removed because they tend to distort the figure by pulling the center towards outliers. The map shows the Anglo-American countries on the right side of the figure as similar in their collaboration patterns. In this projection, the Asian nations are positioned towards the bottom-left side—with the exception of Japan—with the nations of the Middle East as a nearby cluster.

Continental Europe is in the middle. The European position is caused by the dense network of collaborations among the core EU nations (such as France, the Benelux countries, and Germany). Portfolios of EU nations are influenced by the funding of the European Commission's science and cohesion policies requiring collaboration. Japan is not visible on this map because its node is hidden behind France in the center area; the node and label for Japan can be made visible by choosing the (alternative) "label view" in VOSViewer. Certain other nations such as Argentina, Brazil, and Mexico are also related to this set, whereas Chile, for example, is more exclusively related to Spain. The somewhat specific positions of Italy and Austria at the peripheries of this map are noteworthy showing that the accession countries of Central and Eastern Europe are integrated in a triangle involving these two nations and Germany.

*c. Center and periphery in the network*

Figure 3 shows the network among 42 nations forming a strong component in the network of international coauthorship relations in 2011.

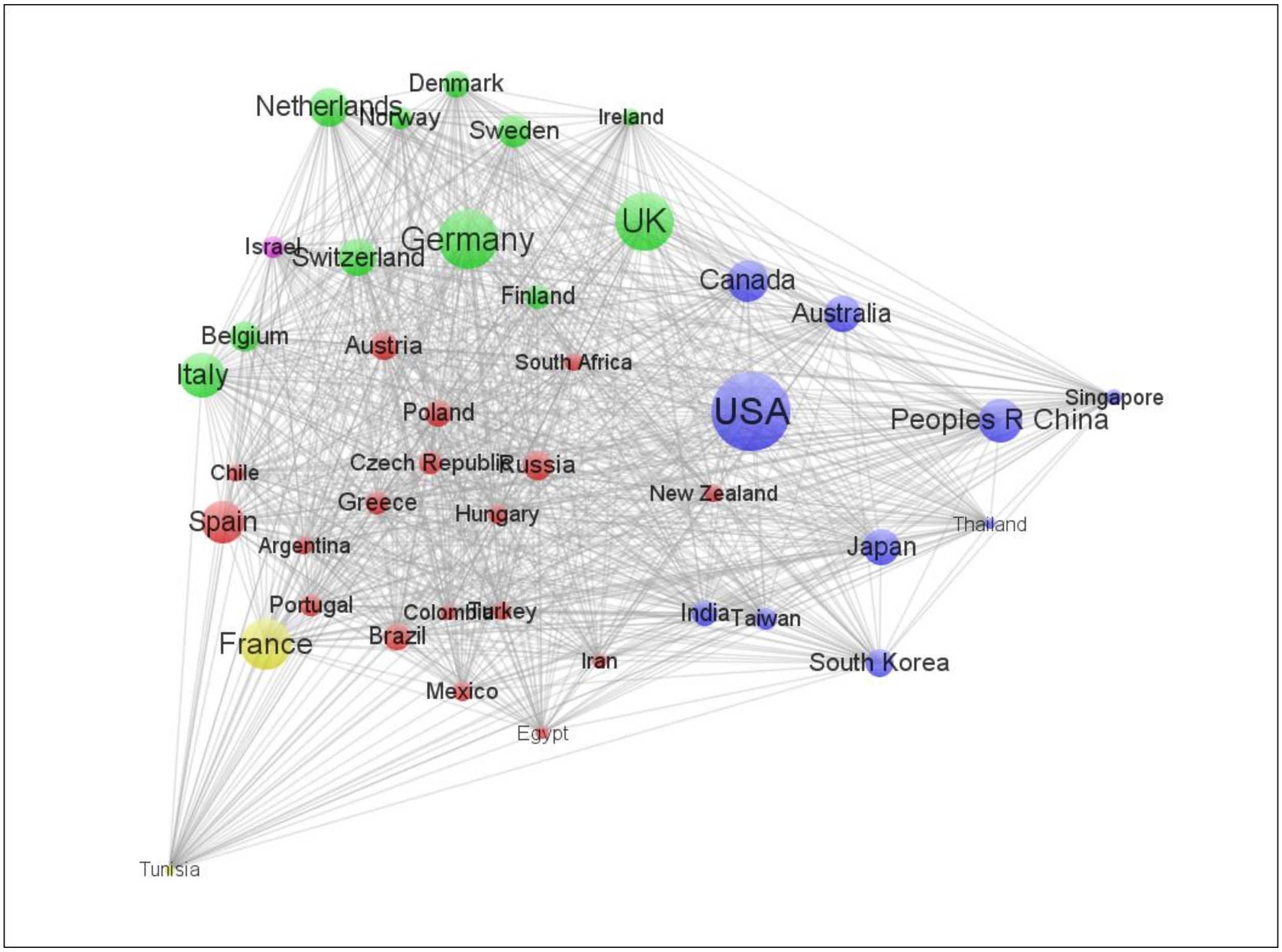

**Figure 3**: The strong component of 42 nations in the center of the network (no normalization implied). Nodes are normalized in terms of their numbers of relations (i.e., degree distributions); VOSViewer used for clustering, coloring, and mapping. Available at http://www.vosviewer.com/vosviewer.php?map=http://www.leydesdorff.net/intcoll/core42map.txt&network=http://www.leydesdorff.net/intcoll/core42net.txt&n_lines=3000&label_size=1.35 .

This figure shows the major players in the network in terms of international coauthorship relations. In contrast to the ranking of shares of publications in terms of addresses—China is also second behind the USA in terms of fractional counts—this figure shows, among other things, that China is not (yet) so active in terms of international coauthorship as are advanced industrial countries (e.g., the UK and Germany; National Science Board, 2012, at p. 5-37; cf. Plume, 2011). However, in contrast to data examined in 2005/2006, China is now part of the central group.

The polar position of France (at the bottom left) is noteworthy and can be considered as a consequence of its leading position (along with Spain) in collaborations with Mediterranean and Romance-language-speaking countries. Despite the nearly global use of English as the language of research publication (99.1% in this data), there are still distinct collaborative groupings of Francophone countries in Africa (Adams, King, and Hook, 2010; Adams *et al.*, in preparation) and Luso-/Hispanophone nations in central and South America. These networks point to cultural and economic factors underlying regional differentiation in the global patterns.

*d. The international environments of nations*

As noted, individual nations may not be visible on the global map at http://www.leydesdorff.net/intcoll/intcoll.htm because of insufficient representation with regard to thresholds. Regional analyses, with more relaxed thresholds on volume of activity and collaboration, enable the user to extend this analysis and show how countries may become local hubs to emerging regional networks (Adams, King, and Hook, 2010; Adams *et al.*, 2011).

Indonesia, for example, has 559 papers in the set, but fractionally counted these add up to only 227.9 coauthored documents. Using Pajek (or any other network analysis program), the user can bring the co-authorship neighborhood of a specific nation to the fore, as in Figure 4 for Indonesia: 86 countries are included in this so-called ego-network, but with (sometimes single) co-authorship relations. [4]

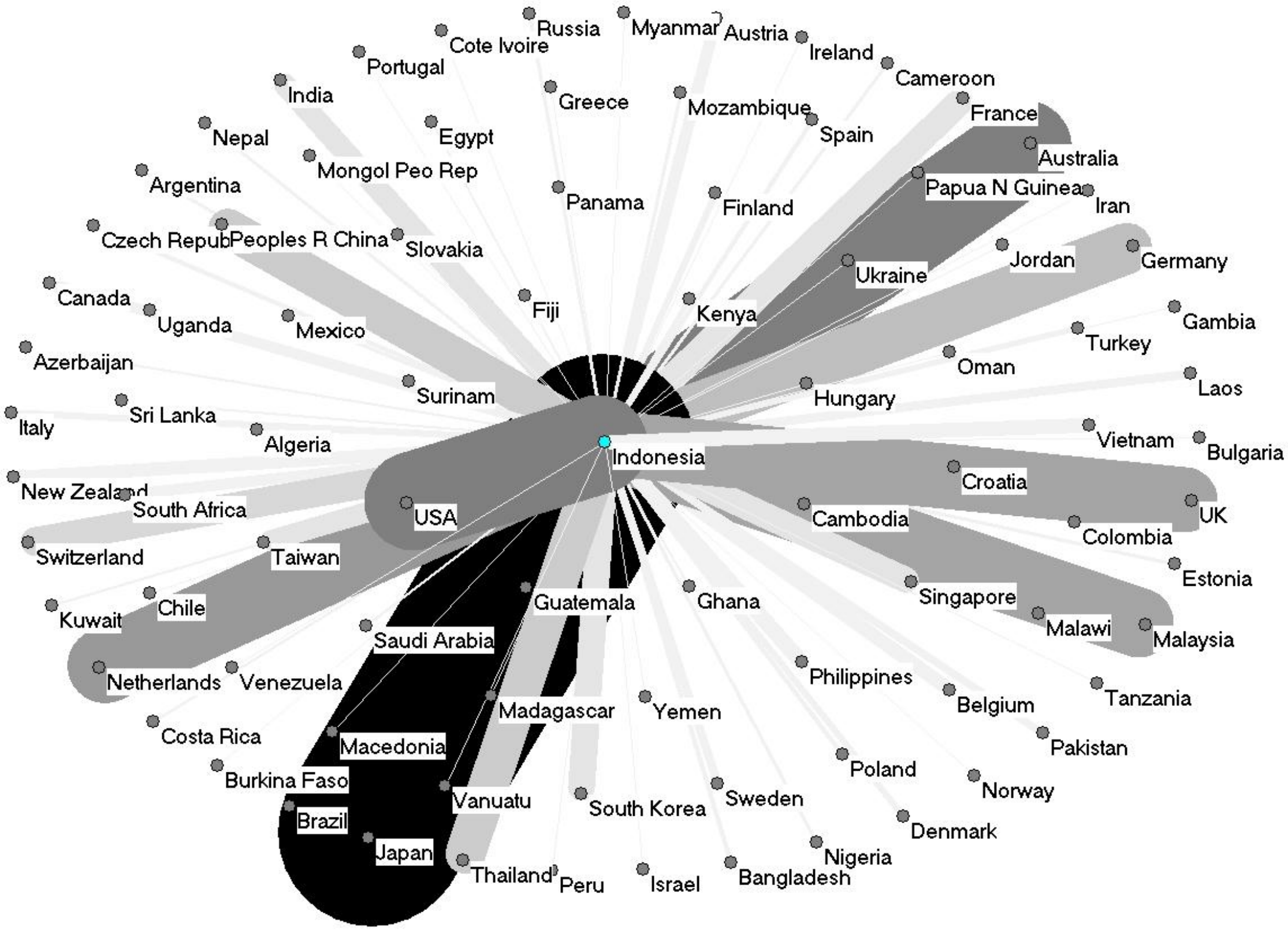

**Figure 4**: 1021 international coauthorship relations with authors in 86 other countries on the basis of 559 documents with an Indonesian address in 2011; *k*=1 network in Pajek. An equivalent file can be webstarted in VOSViewer using

[4] The file for Indonesia is brought online for didactic purposes at http://www.leydesdorff.net/intcoll/indonesia.paj. The subsequent steps after opening the file in Pajek are as follows:

1. Read the full network ("coocc201.net"; included in the file "indonesia.paj");
2. Network > Partition > *k*-neighbours; select node number and distance 1.
3. Operations > Network + Partition > extract subnetwork 0-1; "0" for ego, "1" for *k*=1 neighbours;
4. Partition > Make Cluster > 1 (only *k*=1 neighbours)
5. Operations > Network + Partition > Transform > Remove Lines > Inside Cluster 1 (that is, links among k-neighbours)
6. Draw > Network + first partition
7. You may have to turn on labeling not only the cluster under Options in the draw screen; otherwise one only sees the k-neighbours labeled.

http://www.vosviewer.com/vosviewer.php?map=http://www.leydesdorff.net/intcoll/indon_map.txt&network=http://www.leydesdorff.net/intcoll/indon_net.txt&n_lines=3000&label_size=1.35 .

Although not a major player in the global science system, Indonesia is strongly networked to the extent that on average each paper is coauthored 2.5 times (= 559 / 227.9). The main international relations are with advanced industrial neighbors in the Asian-Pacific region, the USA, and specific European nations. Many of these relations may be a consequence of scholars having studied abroad as postdocs or Ph.D. students.

Given the origin and readership of this journal, we were asked to pay additional attention to Latin America, Spain, and Portugal. Figure 5 provides the collaboration network among these nations including some which can be considered francophone (e.g., Haiti) or anglophone (e.g., Trinidad Tobago), but which one can expect to be integrated in the region.

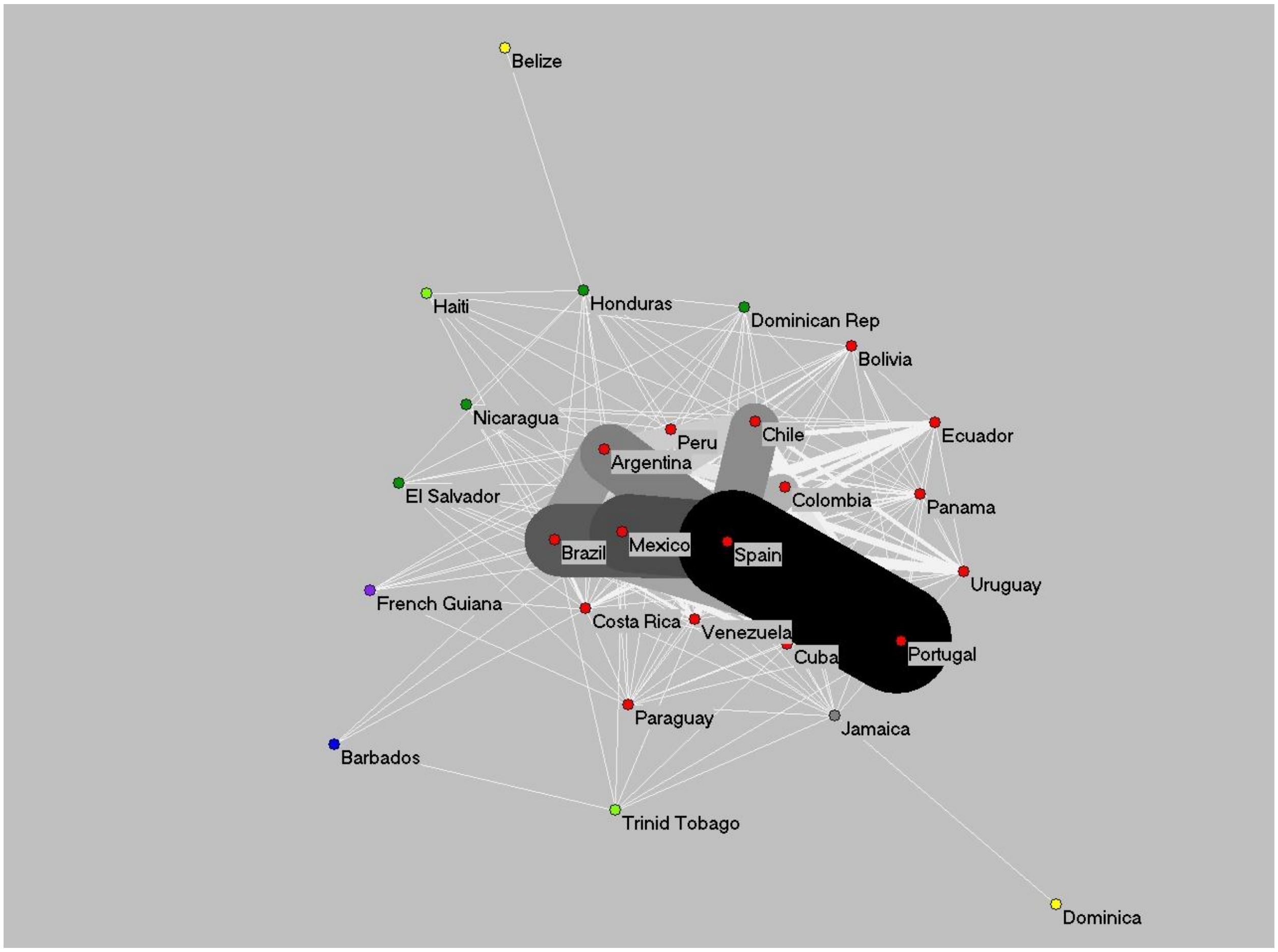

**Figure 5**: Coauthorship network of 27 nations with relevance for Latin America.

Figure 5 first shows the much stronger connection between Spain and Portugal—as both EU member states—when compared with the linguistic relations overseas. Spain has remained a hub between the EU and Latin America more than Portugal (Glänzel *et al.*, 2006). Relations among Chile, Brazil and Argentina are less developed than those between each of these countries and Spain (Presmanes & Zumelzu, 2003). Countries with languages other than Spanish or Portuguese are peripheral to this network as are some nations in central America. In summary, south-south collaboration remains peripheral when evaluated from the global perspective (Adams *et al.*, in preparation).

## Summary and conclusions

The network of coauthorship relations offers a perspective on the ranks and positions of countries which provides an alternative to ranking shares of publications and citations. The core group of collaborating nations is dominated by a subset of research-intensive Western-European nations and the USA. This configuration was challenged during the 1990s and early 2000s by the arrival of new entrants at the global level. As the analysis shows, all the nations of the world are now participating in this process of globalization. Whereas Leydesdorff & Wagner (2008) once feared that a small set of (approximately 14) nations could monopolize the network by reproducing historical patterns, the leading group has tripled to more than 40 nations (Figure 3) in the last five years, suggesting a different dynamic operating at the global level. Thus, the development is more inclusive than before, with features more similar to an open system with some regional differentiation than the core-periphery grouping that characterized the global system in the past.

The globalization of co-authorship relations at current levels—with almost 25% of the relevant papers internationally coauthored, but carrying almost 40% of the institutional addresses in the file—can be expected to have changed (or reflect changes in) the structure of science and the dynamics of knowledge creation in the core set. Whereas the context of discovery for generating knowledge claims was previously considered mainly a social context while the context of validation was envisioned at the global (or universal) level (Popper, [1935] 1959), nowadays the two contexts are increasingly intermingled. Gibbons *et al.* (1994) hypothesized a third "context of application" that allows stakeholders to participate in the process of knowledge production and validation (cf. Lepori, 2011). National science policies based in institutions created in the 20th century may be less effective in influencing such a complex and adaptive system developing at the global level.

**Acknowledgement**

We thank Thomson Reuters for access to the data; some of us acknowledge support from the SSK (Social Science Korea) Program funded by National Research Foundation of South Korea; NRF-2010-330-B00232.

## References

Adams, J. (2012). Collaborations: the rise of research networks. *Nature, 490*, 335-336.

Adams, J., Gurney, K., Hook, D., & Leydesdorff, L. (in preparation). International collaboration clusters in Africa.

Adams, J., King, C. & Hook, D. (2010). *Global Research Report: Africa*. Leeds, UK: Evidence Thomson Reuters. ISBN 1-904431-25-9

Adams, J., King, D., Pendlebury, D., Hook, D. & Wilsdon, J. (2011). *Exploring the changing landscape of Arabian, Persian and Turkish researc*h. Leeds, UK: Evidence Thomson Reuters. ISBN 1-904431-27-5

Ahlgren, P., Jarneving, B., & Rousseau, R. (2003). Requirements for a Cocitation Similarity Measure, with Special Reference to Pearson's Correlation Coefficient. *Journal of the American Society for Information Science and Technology, 54*(6), 550-560.

Barjak, F., Eccles, K., Meyer, E. T., Robinson, S., & Schroeder, R. (2013). The Emerging Governance of E-Infrastructure. *Journal of Computer-Mediated Communication*.

Ben-David, J. (1971). *The Scientist's Role in Society: A comparative study*. Englewood Cliffs, NJ: Prentice-Hall.

Choi, S. (2012). Core-periphery, new clusters, or rising stars?: international scientific collaboration among 'advanced' countries in the era of globalization. *Scientometrics, 90*(1), 25-41.

Crane, D. (1972). *Invisible Colleges*. Chicago: University of Chicago Press.

Gibbons, M., Limoges, C., Nowotny, H., Schwartzman, S., Scott, P., & Trow, M. (1994). *The new production of knowledge: the dynamics of science and research in contemporary societies*. London: Sage.

Glänzel, W., & Schlemmer, B. (2007). National research profiles in a changing Europe (1983–2003) An exploratory study of sectoral characteristics in the Triple Helix. *Scientometrics, 70*(2), 267-275.

Glänzel, W., Leta, J., & Thijs, B. (2006). Science in Brazil. Part 1: A macro-level comparative study. *Scientometrics, 67*(1), 67-86.

Katz, J. S., & Martin, B. R. (1997). What is research collaboration? *Research Policy, 26*(1), 1-18.

Kwon, K. S., Park, H. W., So, M., & Leydesdorff, L. (2012). Has Globalization Strengthened South Korea's National Research System? National and International Dynamics of the Triple Helix of Scientific Co-authorship Relationships in South Korea. *Scientometrics, 90*(1), 163-175. doi: 10.1007/s11192-011-0512-9
Larivière, V., Gingras, Y., & Archambault, É. (2006). Canadian collaboration networks: A comparative analysis of the natural sciences, social sciences and the humanities. *Scientometrics, 68*(3), 519-533.
Laudel, G. (2002). What do we measure by co-authorships? *Research Evaluation, 11*(1), 3-15.
Lepori, B. (2011). Coordination modes in public funding systems. *Research Policy, 40*(3), 355-367.
Leydesdorff, L., & Sun, Y. (2009). National and International Dimensions of the Triple Helix in Japan: University-Industry-Government versus International Co-Authorship Relations. *Journal of the American Society for Information Science and Technology 60*(4), 778-788.
Leydesdorff, L., & Wagner, C. S. (2008). International collaboration in science and the formation of a core group. *Journal of Informetrics, 2*(4), 317-325.
Melin, G., & Persson, O. (1996). Studying research collaboration using co-authorships. *Scientometrics, 36*(3), 363-377.
National Science Board. (2012). *Science and Engineering Indicators*. Washington DC: National Science Foundation; available at http://www.nsf.gov/statistics/seind12/.
Persson, O. (2010). Are highly cited papers more international? *Scientometrics, 83*(2), 397-401.
Persson, O., Glänzel, W., & Danell, R. (2004). Inflationary bibliometric values: The role of scientific collaboration and the need for relative indicators in evaluative studies. *Scientometrics, 60*(3), 421-432.
Plume, A. (2011). Tipping the balance: The rise of China as a science superpower. *Research Trends*(22), at http://www.researchtrends.com/issue22-march-2011/tipping-the-balance-the-rise-of-china-as-a-science-superpower/.
Popper, K. R. ([1935] 1959). *The Logic of Scientific Discovery*. London: Hutchinson.
Presmanes, B., & Zumelzu, E. (2003). Scientific cooperation between Chile and Spain: Joint mainstream publications (1991-2000). *Scientometrics, 58*(3), 547-558.
Schott, T. (1991). The World Scientific Community: Globality and Globalisation,. *Minerva 29*, 440-462.
Wagner, C. S. (2008). *The New Invisible College*. Washington, DC: Brookings Press.
Wagner, C. S., & Leydesdorff, L. (2005). Network Structure, Self-Organization and the Growth of International Collaboration in Science. *Research Policy, 34*(10), 1608-1618.
Wagner, C. S., & Wong, S. K. (2012). Unseen science? Representation of BRICs in global science. *Scientometrics, 90*(3), 1001-1013.
Wagner, C. S., Park, H., Adams, J., & Leydesdorff, L., (in preparation). Policy Implications of the Global Network of Science.
Woolgar, S. W. (1976). The identification and definition of scientific collectivities. In G. Lemaine & e. al. (Eds.), *Perspectives on the Emergence of Scientific Disciplines* (pp. 233-245). The Hague: Mouton.
Zhou, P., & Glänzel, W. (2010). In-depth analysis on China's international cooperation in science. *Scientometrics, 82*(3), 597-612.